\documentclass[10pt]{article}

\usepackage{epsf}
\usepackage{amsmath}
\usepackage{tabularx}

\newcommand{\qaqs}{
\setlength{\unitlength}{0.5mm}
\raisebox{-4.2mm}{
\begin{picture}(40,20)(-15,-10)
\put(-19,10){\makebox(0,0){$\bar q$}}
\put(-19,-10){\makebox(0,0){$q$}}
\put(19,-10){\makebox(0,0){$q$}}
\put(19,10){\makebox(0,0){$\bar q$}}
\put(-15,-10){\vector(3,2){9}}
\put(-7.5,-5){\line(3,2){18}}
\put(15,10){\vector(-3,-2){9}}
\put(-15,10){\line(3,-2){9}}
\put(0,0){\vector(-3,2){9}}
\put(0,0){\vector(3,-2){9}}
\put(15,-10){\line(-3,2){9}}
\put(0,0){\circle*{4}}
\end{picture}}}       

\newcommand{\qqs}{
\setlength{\unitlength}{0.5mm}
\raisebox{-4.2mm}
{\begin{picture}(40,20)(-15,-10)
\put(-19,-10){\makebox(0,0){$q$}}
\put(-19,10) {\makebox(0,0){$q$}}
\put(19,-10) {\makebox(0,0){$\bar q$}}
\put(19,10)  {\makebox(0,0){$\bar q$}}
\put(-15,-10){\vector(3,2){9}}
\put(-7.5,-5){\vector(3,2){18}}
\put(7.5,5)  {\line(3,2){9}}
\put(-15,10) {\vector(3,-2){9}}
\put(-7.5,5) {\vector(3,-2){18}}
\put(7.5,-5){\line(3,-2){9}}
\put(0,0){\circle*{4}}
\end{picture}}}         

\newcommand{\tadsigma}
{\setlength{\unitlength}{1.5 mm}
\raisebox{- 3 mm}
{\begin{picture}(14,10)(-7,-5)
\put(-7,-5){\line(1,0){14}}
\multiput(0,-5)(0,1.0){4}{\line(0,1){0.75}}
\put(0,-5){\circle*{0.8}}
\put(0,-0.8) {\circle*{1.5}}
\put(1.8,-2.5){\makebox(0,0){$\sigma'$}}
\end{picture}}}

\newcommand{\hsloop}
{\setlength{\unitlength}{1.5 mm}
\raisebox{-5 mm}
{\begin{picture}(14,10)(-7,-6.3)
\put(-7,-5){\line(1,0){14}}
\multiput(0,-5)(0,1.0){4}{\line(0,1){0.75}}
\put(0,-5){\circle*{0.8}}
\put(0,-0.8) {\circle*{0.8}}
\put(0,2.1) {\circle{6}}
\put(1.8,-2.5){\makebox(0,0){$\sigma'$}}
\end{picture}}}

\title{\bf On the Nucleon Stability and $\Sigma_N$ Term Puzzles} 
\author{S. Ying\thanks{e-mail: sqying@fudan.edu.cn}\\
         Research Center for Theoretical Physics, Physics Department\\ 
         Fudan University, Shanghai 200433, China}

\begin{document}
\maketitle
\begin{abstract}

  A stable soliton configuration for a nucleon emerges when the
nucleon stability and $\Sigma_N$ term discrepancy problems are studied
semi-quantitatively within a local theory developed recently. The
approach developed here goes beyond the mean field or Hartree--Fock
approximation by taking into account of the non-perturbative wave
function renormalization in a way that is consistent with the chiral
Ward--Takahashi identities of QCD.  The stability condition for a
nucleon and the recently extracted nucleon $\Sigma_N$ term are used to
estimate the radius $R$ of the soliton. It is found that 0.67 fm $\le
R \le$ 0.78 fm. A new mechanism for the stability of a nucleon is
proposed. The discrepancy between the nucleon $\Sigma_N$ term
extracted from pion--nucleon scattering data and the one from baryonic
spectra is resolved by assuming the existence of a metastable virtual
color superconducting phase for the strong interaction vacuum. Under
such a scenario, the difference between the energy density of the
chiral symmetry breaking phase and the metastable color
superconducting phase is found to be 0.41 GeV/fm$^3$ $\le
\Delta\epsilon\le $ 0.85 GeV/fm$^3$.
\end{abstract}

\section{Introduction}

Two puzzles for a nucleon, which is believed to connect to its
chiral properties, remain to be explained theoretically. The first one
concerns with the mechanism for the stability of a nucleon; the second
one is the nucleon $\Sigma_N$ term discrepancy problem, which is made
more severe by recent analysis of the pion--nuclon scattering data.

1) Despite the success of the bag and/or non-topological solitonic
type of models for a nucleon in the case of a single nucleon, a
serious potential problem for these models exists when one tries to
describe many nucleon systems, like a nucleus, base on these models.
On the one hand, there are empirical evidences which imply that
a nucleon inside of a nucleus keeps most of its identity.  On the
other hand, the theoretical models allow the solution for a nucleon
inside a nucleus to change significantly compared to the free space
solution to favor a bag or a soliton of multi-quarks rather than the
phenomenologically successful picture of a nucleus that can be
described using the nucleon and meson degrees of freedom.  Albeit
arguments for a kind of equivalence between these two pictures were
put forward in literature, the quantitative justification of this
picture in realistic situations is still lacking.  It implies that
perhaps certain elements of an essential physics have not been
understood so far. It is expected that a study of this possibility
using the local theory recently developed could lead to a way out of
the dilemma \cite{tp1,tp2,tp3,tp4}.

2) Due to the approximate chiral symmetry in strong interaction, which
is only explicitly broken by a tiny current mass for the light quarks,
the low energy dynamics involving pion is very accurately constrained
by the chiral symmetry of QCD. Indeed the Goldberger--Treiman
relation, the Adler--Weisberger relation and other chiral relations
are confirmed by data \cite{recent,gp-exp} even with an extension of
the PCAC relation \cite{pcac1,pcac2}, which includes the possibility
that there is a vector type virtual color superconducting phase.  Such
an expected success fails to reappear in the next leading order in the
current quark mass. The resulting nucleon $\sigma_N$ term, which is a
measure of the scalar charge of the nucleon multiplied by the current
quark mass, seems to be poorly connected to the on shell pion nucleon
compton scattering data. The analysis of the most current data
\cite{Olsson,GWash} generates even greater values for $\Sigma_N$ on
the Cheng--Dashen point than the old one \cite{Gasser}--making the
descrepancy larger ($\sim 100\%$ effect).  The lattice QCD evaluation
\cite{latt1,latt2} of the same quantity also produces a larger value.
Under the conventional chiral symmetry breaking picture, this is
indeed very puzzling: why the leading order relations are so
accurately confirmed by data and at the same time the next leading
order one fails $100 \%$, given the fact that the expected error
should be at most around $m_\pi/M_A \sim 10\%$ with $m_\pi$ the mass
of a pion and $M_A\sim 1$ GeV. Assuming a large strangeness content
for a nucleon may not be consistent with other observables for the
nucleon \cite{Gasser}. The way to resolve the nucleon $\Sigma_N$
problem without certain new insight about the structure of the nucleon
is a challenge if the chiral symmetry is regarded to be only slightly
broken at low energy, as it manifests itself in the leading order and
in many other observables.

The reason for the leading order results not to be affected by whether
or not there is a metastable color superconducting phase
\cite{pcac1,pcac2} is not hard to understand. The quantities that are
sensitive to the possible metastable color superconducting phases are
evaluated on the pion mass shell, resulting in a pion dominance.  The
quantity, namely $g_A$, which is not evaluated on the pion mass shell
is however a quantity that is insensitive to the low energy chiral
dynamics. In fact the scale that determines the change of $g_A$ with
the momentum transfer is controlled by a scale of order $M_A\sim 1$ GeV
\cite{MA}. The nucleon $\Sigma_N$ term on the other hand does not
enjoy such an insensitivity.

It is therefore interesting to study whether or not these difficulties
of the conventional picture for the nucleon could provide a useful
source of information for a semi-quantitative discussion of various
basic properties of a nucleon in the light of the local theory developed
recently.

There are several new features in the local theory. First of all, it
is based on an 8--component ``real'' representation for fermions,
which is not equivalent to the currently adopted frame--work for
relativistic processes at finite density \cite{tp1,tp2,tp3,tp4}. Some of
the advantages of this representation are discussed in
Ref. \cite{ptct}. Second, the statistical gauge invariance, dark
component for local observables and the statistical blocking effects
can be studied in the new frame-work \cite{tp1,tp2,tp3,tp4}. It would be
interesting to study what is its implication on a possible
simultaneous solution of the aforementioned problems.  It serves as a
further consistency check of the mechanisms proposed here and in our
earlier publications.

Qualitative discussion of such a solution to the nucleon stability
problem is proposed recently \cite{tp1,tp2,tp3,tp4}, in which it is argued
that the statistical blocking effects of the strong interaction vacuum
state can be the underlying physical mechanism for the stability of a
nucleon inside of a nucleus and nuclear matter\cite{tp1}. It remains to
be investigated quantitatively as to whether or not such a mechanism 
is right in the sense that 1) it respects constraints of
the chiral Ward--Takahashi identities and 2) it could reproduce at
least two of the most important physical properties of the nucleon,
namely, its size and its mass with reasonable model assumptions and
parameters. This work aims at a semi-quantitative study of this
question. A model picture for the nucleon has to be established and the
physics and order of magnitude estimate of the possible quantitative
properties of the model could then be derived.

The nucleon $\Sigma_N$ term problem is also not a qualitative problem
in the local theory if one assumes that there is at least one
metastable color superconducting phase \cite{pcac1}. New developments
are made both in the theoretical and phenomenological directions
\cite{review}, which indicate that such a possibility can be oberved
and further more is favored by the evidences considered. However these
studies contains very little specific properties of the nucleon
structure. It seems that a more quantitative study of the nucleon
$\Sigma_N$ problem is the natural next step.

  The paper is organized in the following way. Section
\ref{sec:global} contains an introduction of the chiral models for low
and intermediate energy strong interaction, a discussion of the vacuum
phenomenology in terms of wave function renormalized quasi-particles
of the spontaneous chiral symmetry breaking, an introduction of a
model for a nucleon and a determination of its size under such a
scenario when the value of $\Sigma_N$ is kept constant. A
semi-quantitative mechanism for the stability of a nucleon inside of a
nucleus and/or nuclear matter is suggested and discussed in Section
\ref{sec:stability}. It is based upon the local theory in which the
fermion fields are represented by an 8-component ``real'' spinor. The
possible role played by the metastable color superconducting phases
for the strong interaction vacuum state is discussed in section
\ref{sec:local}. The discrepancy of the nucleon $\Sigma_N$ term
extracted from different sources is resolved by assuming the existence
of a virtual color superconducting phase for the strong interaction
vacuum state, in which, a virtual color superconducting component for
a nucleon lives. The difference in energy density between the true
chiral symmetry breaking ground state and the virtual color
superconducting phase is estimated based on available data. The
section \ref{sec:discussion} contains further discussions.  A summary
is given in the last section.

\section{The Global $\Sigma_N$ Term and a Model for Nucleons}
\label{sec:global}

The standard model for strong interaction is QCD from which relevant
non-perturbative information about the vacuum and hadron structure
interested in this study is not easy to extract. Although lattice
simulation could provide a first principle investigation of the
problem, it is still not satisfactory. Full QCD is also too
complicated for the present purpose. Simplified models constructed
based on the slightly (explicit) broken chiral symmetry of the QCD
Lagrangian in the fermionic sectors are frequently used, which are
regarded as corresponding to the effective Lagrangian of QCD when the
gluonic degrees of freedom are functionally integrated out. This
effective Lagrangian is further simplified by assuming that only
4-point contact 4-point quark--quark interaction terms are
dominant. The well known 't Hooft interaction due to instanton
\cite{Instn} is one of them.  This kind of models are quite
successful in describing a large collection of the hadronic phenomena
(see, e.g., Refs. \cite{weise,Klev,Hats,Glzman,Ivanov} and the references
therein). It indicates that these models contain elements of the truth
about the physics of strong interaction.

In the present section, attempts are made in the following to justify
the picture that nucleons are made up of loosely bound constituent
quarks in a chiral bag.  The mass and wave function renormalization of
these dressed quarks are generated by the spontaneous breaking of the
chiral symmetry, which is governed by an 4--point quark--quark
interaction. Albeit the long range confinement effects are not
considered in these models, they are expected to be reasonable ones
for the ground state chiral properties of hadrons, in which quarks are
not separated far apart. The underlying QCD plays very little direct
role as far as the chiral properties are concerned once these
quantities are renormalized using relevant data from
observations. Other quantities of interest here that detailed QCD
dynamics may play more important role like the confinement, the wave
function renormalization of the quark fields, the bag constant and the
energy density of the possible virtual color superconducting phase
are, instead of predicted by the models, fitted to experimental
data. If possible, a comparison with available lattice QCD data is
going to be carried out to test the consistency of the model
assumptions.

\subsection{The strong interaction models}

  The 4--fermion interaction models for the light quark system
involving up and down quarks have the generic form
\begin{eqnarray}
  {\cal L} &=&  \overline\psi  \left (i\rlap\slash\partial-m_0
       \right ) \psi + {\cal L}_{int}
\label{Lagrangian}
\end{eqnarray}
with the 4--fermion interaction terms classified into two categories
in the quark--antiquark channel
\begin{eqnarray}
{{\cal L}}_{int} & = & \left. \qaqs \right\}\begin{array}{c}
color\\singlet\end{array}+\left. \qaqs \right\}
\begin{array}{c}
     color\\octet
\end{array} +
Fierz\hskip 0.08in
		 term\nonumber\\
	   \nonumber\\
	   & = & {{\cal L}}^{(0)}_{int} +
{{\cal L}}^{(8)}_{int}.\label{LAG1}
\end{eqnarray}
${{\cal L}}^{(0)}_{int}$ generates quark--antiquark scattering in
color singlet channel and ${{\cal L}}^{(8)}_{int}$ generates
quark--antiquark scattering in color octet channel.  For ${{\cal
L}}^{(0)}_{int}$, the well known two flavor chiral symmetric Nambu
Jona--Lasinio (NJL)  interaction can be chosen, namely
\begin{eqnarray}
{{\cal L}}^{(0)}_{int} & = & G_0 \left [ (\overline\psi\psi)^2 +
(\overline\psi i\gamma^5\mbox{\boldmath{$\tau$}}\psi)^2 \right ].
\label{NJLL}
\end{eqnarray}
The color octet ${\cal L}^{(8)}$ is written in the quark--quark
(antiquark--antiquark) channel form for our purposes, namely
\begin{eqnarray}
{{\cal L}}^{(8)}_{int} & = & \left. \qqs \right
\}\begin{array}{c}
				 color \\
				 triplet \end{array} +
\left. \qqs\right\}
				 \begin{array}{c}
				 color \\ sextet
				 \end{array}
		  + (q\leftrightarrow \overline q)\nonumber\\
		  & = &{{\cal L}}^{(3)}_{int} +
{{\cal L}}^{(6)}_{int}.\label{lag8}
\end{eqnarray}
The color sextet term is repulsive in the one gluon exchange case.
Due to the non-existence of colored baryon containing three quarks in
nature, it is assumed to be generally true. So we restrict ourselves
to the attractive color triplet two quark interaction terms. The
attractive color triplet quark bilinear terms can be classified
according to their transformation properties under Lorentz and chiral
$SU(2)_L\times SU(2)_R$ groups \cite{Ann1}. In general, if only terms
without derivative in fermion fields are considered, ${\cal
L}_{int}^{(3)}$ has the following form
\begin{eqnarray}
{{\cal L}}^{(3)}_{int} &=&\frac 1 2 \sum_r G_r\sum_{ab} C^r_{ab}
(\overline\psi
       \Gamma_a^r\widetilde{\overline\psi})(\widetilde\psi\Gamma^r_b\psi),\label{
L3}
\end{eqnarray}
with $\Gamma^r_a$, $\Gamma^r_b$ matrices in Dirac, flavor and color
spaces generating representation ``$r$''.  The tilded fermion field
operators $\widetilde\psi$ and $\widetilde{\overline\psi}$ in the
4-component representation are defined as
\begin{eqnarray}
\widetilde\psi & = & \psi^T(-i\tau_2)C^{-1},\\
\widetilde{\overline\psi} & = & Ci\tau_2\overline\psi^T.
\end{eqnarray}
In the 8-component representation for the fermion field in the local
theory, these two identities are implemented as a pair of constraints.

 Operator $\widetilde\psi\Gamma^r_b\psi$ belongs to an irreducible
representation ``r'' of chiral, Lorentz and color groups and
$\overline\psi\Gamma^r_a\widetilde{\overline\psi}$ belongs to the
conjugate representation.  Coefficients $C_{ab}^r$ render the
summation $\sum_{ab}\ldots$ invariant under Lorentz, chiral
$SU(2)_L\times SU(2)_R$ and color $SU(3)_c$ groups. $\{G_r\}$ is a set
of independent 4--fermion coupling constants characterizing the color
triplet quark--quark interactions.

Since the QCD Lagrangian at low energy or long distance is presently
unknown, the coupling constants $G_r$ are not known. Albeit instanton
induced 't Hooft interaction is frequently used in literature, which
can reduce the independent coupling constant to only one, it will not
be assumed here from the start. Rather, by studying the virtual phases
of the strong interaction vacuum states using experimental
observables, some of the important $G_r$, which determines the
neighborhood of the strong interaction vacuum state, can at least be
inferred. The question of whether or not instanton induced interaction
dominates is left to be determined by observations.

For that purpose, two models that have different virtual
superconducting phases for the vacuum state are studied. The half
bosonized version of them with the quark field represented
in an 8-component real representation \cite{tp1,Ann1} are:

\subsubsection{A model for scalar color superconductivity}
\begin{eqnarray}
 {\cal L}_1 & = & \frac 1 2 \overline \Psi\left [i{\rlap\slash\partial}-\sigma-
             i\mbox{\boldmath{$\pi$}}\cdot 
             \mbox{\boldmath{$\tau$}}\gamma^5 O_3-\gamma^5 {\cal
             A}_c\chi^c O_{(+)}-\gamma^5 {\cal A}^c\overline\chi_c O_{(-)}
             \right ]\Psi \nonumber \\
            &   &  -\frac 1{4 G_0} (\sigma^2 +
             \mbox{\boldmath{$\pi$}}^2) 
            + \frac 1  {2 G_{3s}} \overline\chi_c \chi^c,
\label{Model-L-1}
\end{eqnarray}
where $\sigma$, $\mbox{\boldmath{$\pi$}}$, $\overline\chi_c$ and
$\chi^c$ are auxiliary fields, $(\chi^c)^{\dagger} = -
\overline\chi_c$, $G_0$ and $G_{3s}$ are coupling constants of the
model. The matrices $O_i$ (i=1,2,3) are Pauli matrices acting on the
upper and lower 4-components of $\Psi$ and ${\cal A}_c$ (c=R,B,G) are
a set of three $2\times2$ antisymmetric matrices acting on the color
indices of $\Psi$.

 This model Lagrangian \cite{sdiq,tp1} has a color superconducting
phase when the coupling constant $G_{3s}$ is sufficiently large.  When
the phase in which the normal chiral symmetry breaking is the true
ground state of the vacuum, the corresponding color superconducting
phase is a metastable virtual phase \cite{sdiq,tp1} in which the order
parameter $\sigma$ for the normal chiral symmetry breaking phase
vanishes \cite{review}.

\subsubsection{A model for vector color superconductivity}
\begin{eqnarray}
{{\cal L}}_2&=& \frac 1 2 \overline\Psi \left [ i{\rlap\slash\partial} -
\sigma -
 i\mbox{\boldmath{$\pi$}}\cdot\mbox{\boldmath{$\tau$}}\gamma^5 O_3 +
O_{(+)} \left (\phi^c_\mu\gamma^\mu\gamma^5
{\cal A}_c
  +\mbox{\boldmath{$\delta$}}_\mu^c\cdot\mbox{\boldmath{$\tau$}}\gamma^\mu{\cal
A}_c \right ) \right . \nonumber \\ & & \left .- O_{(-)}\left (\overline\phi_{\mu
c}\gamma^\mu\gamma^5{\cal A}^c
  +\mbox{\boldmath{$\overline\delta$}}_{\mu c}\cdot\mbox{\boldmath{$\tau$}}
   \gamma^\mu{\cal
A}^c \right) \right ]\Psi -  \frac 1 {4 G_0}(\sigma^2 
+ \mbox{\boldmath{$\pi$}}^2) \nonumber \\ && - {1\over
2 G_{3v}}
(\overline\phi_{\mu c}\phi^{\mu c} + \mbox{\boldmath{$\overline\delta$}}_{\mu
c}\cdot
\mbox{\boldmath{$\delta$}}^{\mu c}),\label{Model-L-2}
\end{eqnarray}
where $\overline\phi_{\mu c}$, $\phi_\mu^c$ with $(\phi_\mu^\dagger)_c
= - \overline\phi_{\mu c}$, $\mbox{\boldmath{$\overline\delta$}}_{\mu c}$,
$\mbox{\boldmath{$\delta$}}_\mu^c$ with 
$(\mbox{\boldmath{$\delta$}}_\mu^\dagger)_c = -
\mbox{\boldmath{$\overline\delta$}}_{\mu c}$ are auxiliary fields introduced.

 This model Lagrangian \cite{ltb,Ann1,tp1} has a vector color
superconducting phase induced by vector diquark condensation when the
coupling constant $G_{3v}$ is sufficiently large.  The color
superconducting phase also breaks the chiral symmetry.  When the
normal chiral symmetry breaking phase is the true ground state of the
vacuum, the corresponding color superconducting phase is a metastable
virtual phase \cite{ltb,Ann1,tp1} in which the order parameter
$\sigma$ for the normal chiral symmetry breaking phase vanishes 
\cite{review}.

\subsubsection{The normal chiral symmetry breaking phase}

The gap equation for the particles in the normal chiral symmetry
breaking phase for both of these two models can be derived from a
stability equation for the quark mass \cite{Ann1} in the Hartree--Fock
approximation.  The result is
\begin{equation}
  {\sigma^2\over \Lambda^2}\mbox{ln}\left (1+{\Lambda^2\over\sigma^2} 
 \right ) = \left (1-{\pi\over 12\alpha_0 } \right )
\label{gap-eq}
\end{equation}
with 
\begin{equation}
   \alpha_0 = {\widetilde G_0\Lambda^2\over 4\pi},
\end{equation}
where  $\widetilde G_0$ is a linear combination of all the
4--point coupling constants of theory \cite{Ann1}.

Since the vacuum $\sigma_{vac}\approx 350$ MeV, Eq. \ref{gap-eq}
implies $\alpha_0\approx 0.38$ when $\Lambda$ takes the value 0.9 GeV.
The vacuum expectation value $<0|\overline\psi\psi|0>$ that can be
extracted from Gell-Mann, Oakes and Renner \cite{GOR} (GOR) relation
\begin{equation}
   f_\pi^2 m_\pi^2= -m_0 <0|\overline\psi \psi |0>
\label{GORrel}
\end{equation}
is related
to $\sigma_{vac}$ by
\begin{equation}
    \sigma_{vac} = -2 \widetilde G_0 <0|\overline\psi \psi |0>,
\label{mfrel}
\end{equation}
in the mean field or Hartree--Fock approximation.
Here $m_0=(m_u+m_d)/2$, $m_u$ and $m_d$ is the mass of the current
up and down quark respectively.

\begin{figure}[ht]
\begin{center}
\epsfbox{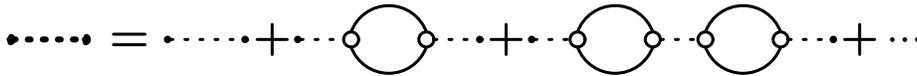}
\end{center}
\caption{\label{Fig:sigprop}\small\em The one loop correction to the $\sigma'$ field
 propagator.}
\end{figure}

For a realistic description of the physics of the strong interaction,
the Hartree--Fock approximation may not be sufficient. One of the most
important corrections is to sum over all the self energy diagrams for
the auxiliary field $\sigma' = \sigma-\sigma_{vac}$, which is shown in
Fig. \ref{Fig:sigprop}. 
The result is a modification of the tree level propagator for
$\sigma'$, which is \cite{Ann1}
\begin{equation}
   D_\sigma(p) = -2iG_0,
\end{equation}
to
\begin{equation}
   D_\sigma(p) = {-2iG_0\over 1-J_n(p)}
\end{equation}
with $J_n \equiv J_n(0)>0$. It causes the anti-screening of the scalar charge
of a dressed quark.

The gap equation is however robust against this kind of corrections
since the same $1/(1-J_n)$ factor enters both of the tadpole terms and
the Hartree part of the loop terms of the stability equation
\cite{Ann1}. So the gap equation in the Hartree approximation with the
dressed $\sigma'$ propagator is of the following form
\begin{equation}
-i\delta\Sigma 
= {1\over 1-J_n} \left ( \tadsigma + \hsloop\right ) = 0,
\end{equation}
where the larger black dot represents the tadpole due to a shift of
the auxiliary field $\sigma$.  The resummation over one-loop
self-energy corrections to the $\sigma'$ field propagator does not
modify of the gap equation derived by using tree level propagator for
the $\sigma'$ field. The Fock terms for the stability equation can
leads to a modification against the naive one loop equations due to
the fact that the dressed propagator for $\sigma'$ inside the loop
depends on momentum and the Fock terms due to exchange of other
auxiliary fields \footnote{The physical Goldstone pion contributions
should be treated using a derivative coupling to the quarks to
implement the chiral symmetry rather than using the pseudo-scalar
coupling of the tree level $\mbox{\boldmath{$\pi$}}$ field 
\cite{lili}. This could
leads to a significant reduction of the pion loop effects, e.g., at
finite temperature \cite{lili}.} does not renormalizes in the same way
as $\sigma'$. This is especially true for the auxiliary fields
corresponding to the ${\cal L}^{(8)}$ term.  The difference should be
ignored in the following discussion based on the large $N_c$
argumentation which suppresses the Folk terms but should be studied
in more detailed researches.

The Hartree--Fock relationship given in Eq. \ref{mfrel}, on the other
hand, has to be modified to
\begin{equation}
    \sigma_{vac} = -{2 \widetilde G_0\over 1-J_n} <0|\overline\psi \psi |0>.
\label{mfrel-1}
\end{equation}
This is because the effective 4-fermion interaction strength
$\widetilde G_0^{eff}= \widetilde G_0/(1-J_n)$ at the level of
approximation considered. Since the resummation over the one-loop
self-energy terms for the $\sigma'$ field does not shift the pole
position of the quark propagator, its effects on the quarks are
represented by a wave function renormalization with $Z_\psi = 1-J_n$,
leading to a reduction of the vacuum expectation value of the
$\overline\psi\psi$ operator against its mean field value.

\subsection{The physical vacuum phenomenology and the global 
            $\Sigma_N$ term}

\subsubsection{The quasiparticle picture for the chiral condensate}

   The physics of the global observables are dominated by the
quasiparticles of the true ground state of the vacuum
\cite{tp1,tp2,tp3,tp4}.  Its physical picture, which could be useful for
the discussions in the following subsection, can be derived from the
gap equation for the dynamical mass of the quarks. The generic form of
the one loop gap equation is of the following form
\begin{equation}
   \sigma_{vac} = i 8 n_f n_c
     \widetilde G_0 \int {d^4p\over (2\pi)^4} {\sigma_{vac} \over p^2-
                             \sigma_{vac}^2 + i \epsilon},
\end{equation}
where $n_f=2$ and $n_c=3$ are the number of flavor and color
respectively. Here, the current
mass $m_0$ of the quarks is assumed to be zero for simplicity. If
treated properly, the solution of this gap equation is beyond the
Hartree--Fock approximation \cite{Ann1}.

In order to project out the contributions of the quasiparticles of the
true ground state of the vacuum, the 4--momentum integration can be
carried out by doing the $p^0$ integration first \cite{tp1,tp3,tp4}.  The
result is
\begin{equation}
   \sigma_{vac} = 4 n_f n_c\widetilde G_0 \int^{\Lambda_3} 
        {d^3 p\over (2\pi)^3} {\sigma_{vac} \over E_p}
\label{sig_mf}
\end{equation}
with $E_p = \sqrt{p^2+\sigma^2}$ the energy of the quasiparticle and
$\Lambda_3$ the formal cut off in the 3-momentum, which is not going
to be used for the numerical evaluations. From Eq. \ref{mfrel-1}, the
scalar density of the vacuum state in the one loop approximation is
\begin{equation}
    <0|\overline\psi \psi |0> = 2 n_f n_c \int^{\Lambda_3} {d^3 p\over (2\pi)^3} 
     \left [-(1-J_n) {\sigma_{vac} \over E_p } \right ]=
     2 n_f n_c \int^{\Lambda_3} {d^3 p\over (2\pi)^3} (1-J_n)\overline u(p) u(p)
\label{sea-schg}
\end{equation}
with $u(p)$ a solution of the Dirac equation for a fermion with mass
$\sigma$ and energy $-E_p$. It can be seen that the
non-perturbative\footnote{ The perturbative wave function
renormalization around the $\sigma_{vac}$ is absent since the
radiative corrections to the self-energy of the quarks is
self-consistently adjusted to zero in the auxiliary field approach
\cite{Ann1} adopted here.}  factor $Z_\psi=1-J_n$ serves as a wave
function renormalization factor for the fermion field $\psi$.

Eq. \ref{sig_mf} implies that the scalar charge density of the vacuum
state is obtained by summing over the scalar charge
\begin{equation}
 \lim_{q_\mu\to 0} j^{(-)}_s(p+q,p) = \overline u(p) u(p) 
\end{equation}
of contributing negative energy fermions in unit space volume
renormalized by factor $1-J_n$.

The physical value for $<0|\overline\psi\psi|0>$ can be computed from model
independent GOR \cite{GOR} relation, which is
\begin{equation}
     <0|\overline\psi\psi |0> = - 0.024 \hspace{2mm} \mbox{GeV}^3
\label{vac-exps}
\end{equation}
when $m_0= (m_u+m_d)/2=7$ MeV is assumed.

The value of $\sigma_{vac}$ is generally taken to be around $0.35$
GeV, which corresponds to the valence quark mass. Assuming
$\Lambda=0.9$ GeV, then Eq. \ref{vac-exps} requires
\begin{equation}
  J_n = 0.21
\end{equation}
which is a positive number as expected. This number is consistent with
the lattice result $Z_\psi = 0.7\sim 0.8$ of Ref. \cite{Wlms}.
 
\subsubsection{Chiral symmetry and pion decay constant}

One of the model independent consequences of the chiral symmetry of
the QCD Lagrangian is derived from the chiral Ward--Takahashi identity
\cite{pcac2}
\begin{equation}
  q^\mu\mbox{\boldmath{$\Gamma$}}^5_\mu = {1\over 2} m_0 D_\pi \gamma^5
        \mbox{\boldmath{$\tau$}} -{1\over 4}\left \{
                \Sigma,\gamma^5\mbox{\boldmath{$\tau$}} \right \}
\label{WTid}
\end{equation}
with $D_\pi$ a scalar function and 
$\mbox{\boldmath{$\Gamma$}}^5_\mu$ the proper axial-vector vertex
for quarks that is dominated by the one pion pole at low $q^2$ and
$\Sigma$ their self-energy. $\mbox{\boldmath{$\tau$}}$ is the set of
Pauli matrices in the isospin space.  It leads to the GOR relation
\cite{pcac1,pcac2} as the first order correction to the chiral
symmetric results for the vacuum case, and, in certain approximate
scheme like the one loop one, an evaluation of the decay constant for
the chiral Goldstone boson \cite{Klev,Ann1}, namely,
\begin{equation}
  f_\pi^2 = {n_f n_c \sigma^2_{vac}\over 8\pi^2} \left ( \ln 
    {\sigma^2_{vac}+\Lambda^2 \over \sigma_{vac}^2}
    - {\Lambda^2\over \sigma_{vac}^2 + \Lambda^2} \right ).
\label{fpi1}
\end{equation}
It is one of the relationships used to fix the model parameters in the
one loop approximation (see, e.g., \cite{Klev}). 

\begin{figure}[ht]
\begin{center}
\epsfbox{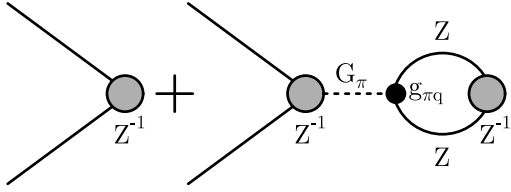}
\end{center}
\caption{\label{Fig:Axial}\small\em The graphical representation of the full
axial-vector vertex function for quarks. Here the grey blobs represent
the corresponding the one pion irreducible proper vertices. The black
dot represents the bare pion--quark coupling $g_{\pi q}$. Only the
non-perturbative renormalization factors due to the self-energy for
the $\sigma'$ field, as indicated in the figure, are taken into
account here.  }
\end{figure}

The picture developed here goes beyond the one loop approximation by
formally summing over the self-energy terms for the auxiliary
$\sigma'$ fields, which results in a wave function renormalization
factor $1-J_n$ for the quasi-particles. Before the spontaneous chiral
symmetry breaking taking place, the chiral Ward--Takahashi identity
requires that there should also be a $1/(1-J_n)$ renormalization
factor for the axial-vector vertex as it is shown in the first graph
of Fig. \ref{Fig:Axial}.  After the chiral symmetry is spontaneously
broken down, a massless Goldstone pion pole appears in the
axial-vector vertex which is shown in the second graph of
Fig. \ref{Fig:Axial}.  It is reasonable to assume that the same
renormalization for the pion--quark coupling vertex. The self-energy
$\Sigma$ that enters Eq. \ref{WTid} is related to the 
quark mass $\sigma_{vac}$ as
\begin{eqnarray}
    \Sigma &\to& {\sigma_{vac}\over 1-J_n}.
\end{eqnarray}
The various renormalization factors are displayed in
Fig. \ref{Fig:Axial}.  for the axial-vector vertex function with the
pion pole separated out.
The renormalized one loop result for the pion decay constant is then
found to be
\begin{equation}
  f_\pi^2 = {n_f n_c \sigma^2_{vac}\over 8\pi^2} \left ( 1-J_n \right )
    \left ( \ln 
    {\sigma^2_{vac}+\Lambda^2 \over \sigma_{vac}^2}
    - {\Lambda^2\over \sigma_{vac}^2 + \Lambda^2} \right ).
\label{fpi}
\end{equation}
The values $\Lambda=0.9$ GeV, $f_\pi=93$ MeV, $\sigma_{vac}\sim 350$
MeV and $J_n = 0.21$ adopted above are consistent with this equation
and thus with the chiral Ward--Takahashi identity \ref{WTid}.

Thus, the approximation approach taken here, which improves the
canonical quasi-particle picture by introducing a non-perturbative
 wave function renormalization factor $1-J_n$, is quit successful
in the vacuum sector.

\subsubsection{Model independent information about the scalar charge 
               (density) 
               inside a nucleon}

In the chiral quark model adopted here, the global nucleon $\Sigma_N$
term, which can be extracted from the low energy pion--nucleon Compton
scattering, can be evaluated in the following way
\begin{equation}
    \Sigma_N(0) = <N|:\overline\psi\psi:|N> = m_0 \left [3 (1-J_n) j_s^{(+)} + 
                               \Delta v 
\delta\hspace{-1mm}<0|\overline\psi\psi |0> \right ].
\label{Sig1}
\end{equation}
The scalar charge for the valence quarks $j_s^{(+)}\approx 1$, $\Delta
v$ is the volume taken by a nucleon and
$\delta\hspace{-1mm}<0|\overline\psi\psi |0>$ is the change of the
scalar charge density of the quarks on the negative energy states
inside of the nucleon.

The global $\delta\hspace{-1mm}<0|\overline\psi\psi |0>$ term
interested here is allowed to differ from zero. It is the sum of all
the renormalized quark scalar charge density of all the contributing
quarks orbits with a distortion of $\sigma_{vac}$ inside of a nucleon
minus those without a distortion of $\sigma_{vac}$. A non-vanishing
value for this quantity thus implies a bag picture for a nucleon with
the scalar field $\sigma$ acting as a medium which is distorted to
form the non-topological ``bag''.

Assuming that $m_0=7$ MeV, and that $\Sigma_N(0) \approx 55\sim 75$
MeV after extracting from the Cheng--Dashen point \cite{Olsson,GWash}
to the $t=0$ point, the vacuum polarization effects are
\begin{equation}
\Delta v \delta\hspace{-1mm}<0|\overline\psi\psi |0> = 5.5 \sim 8.4
\label{del-vac}
\end{equation}
which is relatively large. As it is discussed in the following, it
leads to a non-topological bag picture for a nucleon.

\begin{figure}[ht]
\begin{center}
\epsfbox{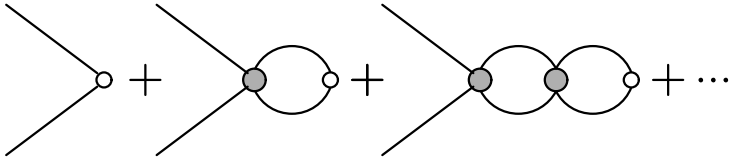}
\end{center}
\caption
{\label{Fig:sclvtx}\small\em The ladder diagrams for scalar charge vertex
function.  The 4--point vertex is shown in Fig. \ref{fig:4ptvtx}.  }
\end{figure}

Before continuing, it should be pointed out that some of the studies of
the nucleon $\Sigma_N$ problem in the 
constituent quark model (see, e.g. Ref. \cite{Hats} and many others)
are based on the following equation
\begin{equation}
    \Sigma_N(0) = <N|:\overline\psi\psi:|N> = m_0 {3 j_s^{(+)}\over 1-J_n}.
\label{oldSig} 
\end{equation}
It can accommodate the old value $\Sigma_N(0)\sim 45$ MeV \cite{Hats}.
Such an equation is obtained by observing that $\Sigma_N(0)$ is a sum
of the static scalar charge of quarks in an additive quark model. The
effect of interaction can be obtained from the scalar vertex function
by summing over ladder diagrams of the form given by
Figs. \ref{Fig:sigprop} and \ref{Fig:sclvtx}, which gives us the
factor $1/(1-J_n)$ in the above equation. The wave function
renormalization of the two amputated quark legs in such an approach is
not taken into account at the same level of approximation as the
vertex function. 

The inclusion of the effects of the wave function
renormalization is required from unitarity point of view. The
interaction of the of the self-energy graphs in Fig. \ref{Fig:sclvtx}
generates poles in the s-channel, which correspond to the physical
mesons \cite{iter}, these new poles should take some of the strength
of the quasi-quark due to unitarity constraint. Without $1-J_n$ term,
the so called ``double or over counting'' of the degrees of freedom is
going to happen.
\begin{figure}[pb]
\begin{center}
\epsfbox{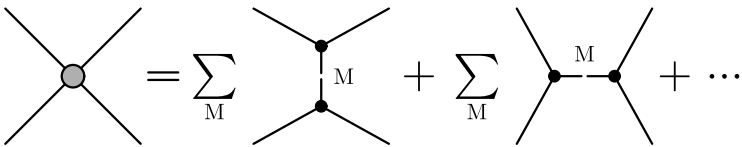}
\end{center}
\caption{\label{fig:4ptvtx}\small\em The 4--point vertex in the scalar channel
in Fig. \ref{fig:scharge}. The summation is over all the auxiliary
fields (M) introduced.  }
\end{figure}
Therefore Eq. \ref{oldSig} is not consistent since it is well known
from the study of the gauge invariance in perturbative quantum
electrodynamics that it is important to consider both the vertex
function renormalization and the wave function renormalization in a
given order in order to maintain the Ward--Takahashi identity. To
sharpen the problem, it should be noticed that Eq. \ref{sea-schg}
should be
\begin{equation}
    <0|\overline\psi \psi |0> = 
     2 n_f n_c \int^{\Lambda_3} {d^3 p\over (2\pi)^3} {\overline u(p)
     u(p)\over 1-J_n}
\label{sea-schg1}
\end{equation}
in such an approximation. It would lead to a significant
inconsistency between Eq. \ref{vac-exps} and the phenomenological
values $\sigma_{vac}=0.35$ GeV, $\alpha_0=0.38$, $\Lambda=0.9$ GeV,
the value of the nucleon $\Sigma_N(0)$ value and the requirement that
$J_n= 0.2\sim 0.3$ from lattice simulation of QCD \cite{Wlms}.

Under Eq. \ref{Sig1}, it is inevitable to include a non-vanishing
value for $\delta\hspace{-1mm}<0|\overline\psi\psi |0>$, as given by
Eq. \ref{del-vac}. It implies a non-topological bag picture for the
nucleon.

The radius of the bag can be determined in the local theory
\cite{tp1,tp2,tp3,tp4} by requiring that a nucleon is stable in a nucleus,
which is revealed in experimental observations. In the aforementioned
local theory, the energy of the lowest energy level in the bag must
be at
\begin{equation}
E=\epsilon_0 - \mu=0.
\end{equation}
This is explained in the next section. Here $\mu$ is the time
component of the statistical gauge field that is related to the
fermion number density through \cite{tp1,tp2}
\begin{equation}
    \rho = {2\over\pi^2} \mu^3
\end{equation}
when the value of the statistical blocking parameter
$\varepsilon_{vac}$ is much smaller than $\mu$ which is non-vanishing
inside of a nucleon. Assuming that $\mu$ is constant inside the bag
and zero outside, its value inside of the bag is related to the bag
radius
\begin{equation}
{2\over \pi^2} \Delta v \mu^3 = {8\over 3\pi} (R\mu)^3 = 3 
\end{equation}
since the fermion number of a nucleon is 3. So
\begin{equation}
   \mu = {3\over 2R} \left ({\pi\over 3} \right )^{1/3}.
\label{muval}
\end{equation}
The energy $\epsilon_0$ can be estimated following the MIT bag model
calculation
\begin{equation}
  \epsilon_0 = {c_0\over R} + \sigma_{in}
\label{eps0}
\end{equation}
with $c_0=2.04$, $\sigma_{in}$ the $\sigma$ field inside of the bag, which is
assumed zero in the authentic MIT bag model. Here, however,
\begin{equation}
       \sigma_{in} = -{2\widetilde G_0\over 1-J_n} \left (<0|\overline\psi\psi|0> + \delta
        \hspace{-1mm}<0|\overline\psi\psi |0> \right ).
\end{equation}
Therefore $\epsilon_0-\mu=0$ implies
\begin{equation}
   \left [c_0-{3\over 2}\left ( {\pi\over 3} \right )^{1/3}   \right ] 
    {1-J_n\over R}
   - {6\alpha_0 \over \Lambda^2} {D\over R^3} = {8\pi\alpha_0\over\Lambda^2}
      <0|\overline\psi \psi |0>,
\end{equation}
where $D=\Delta v \delta\hspace{-1mm}<0|\overline\psi\psi |0> = 5.5 \sim 8.4$ (see
Eq. \ref{del-vac}). 
Substituting the values $\alpha_0=0.38$ and $\Lambda=0.9$
GeV, we have
\begin{equation}
  R= 0.67\sim 0.78\hspace{2mm}\mbox{fm}
\end{equation}
for the core part of a nucleon. This is a quite reasonable range of
numbers.

To get a rough picture for the distribution of $\sigma$ field around a
nucleon, let us estimate $\sigma_{in}$ inside of a nucleon using the
value obtained above
\begin{equation}
\sigma_{in} = -153 \sim -131 \hspace{2mm} \mbox{MeV}
\end{equation}
Therefore we have a bag type of solution for a nucleon in this chiral
model with its basic character determined by the measured global
$\Sigma_N$ term value. The value of the $\sigma$ field outside of the
bag is of order 350 MeV and the value for $\sigma$ inside of the bag
is of order $-153\sim -131$ MeV.

\begin{figure}[ht]
\begin{center}
\epsfbox{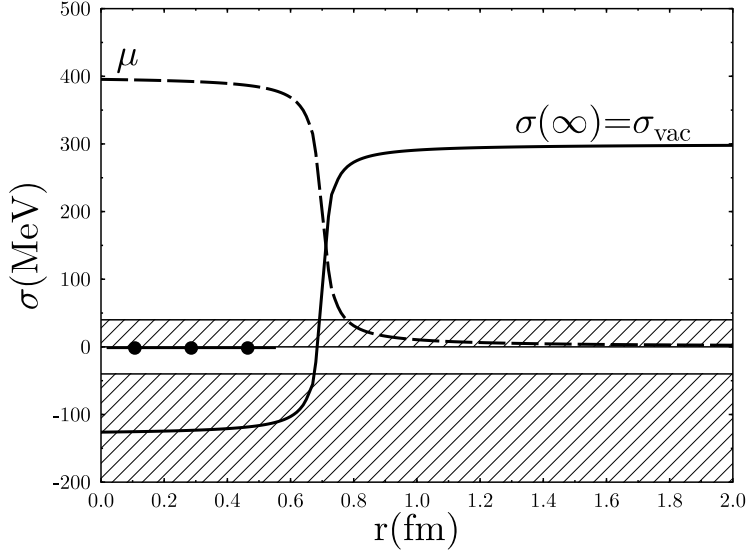}
\end{center}
\caption{\label{Fig:SigShape}\small\em A possible shape for the $\sigma$ field
around a nucleon that can reproduce the global $\Sigma_N(0)$
value. The shaded area contain the forbidden states due to the
statistical blocking effects.  }
\end{figure}

The distribution of $\sigma$ around a nucleon is schematically plotted
in Fig. \ref{Fig:SigShape} in which the three black dots represent the
energy level for the three valence quarks.

\section{The Stability of a Nucleon}
\label{sec:stability}

The stability of a nucleon inside a nucleus is an important enough
problem to the considered, since, as far as it is known now, that the
properties of a nucleus can be described well in terms of the nucleon
degrees of freedom. On the other hand, the current theory tends to
imply a change of the properties of a nucleon inside a nucleon. For
example, the bag constant of a nucleon should be reduced inside of a
heavy nucleus, which would give rise to an increase of the size of the
nucleon. An overlapping configuration is also favored due the
reduction in the surface area between different phases, which is not
usually considered. There is little room for nucleons inside a heavy
nucleus to increase their size before touch each other. There does not
seems to be a reasonable scheme or mechanism to reconcile these two
contradictory pictures at the present.

 The stability problem of a nucleon inside of a nucleus is discussed
qualitatively in Ref. \cite{tp1} under the local theory for the
relativistic finite density problems. The essential mechanism is
derived from the statistical blocking effects in the normal chiral
symmetry breaking phase of the strong interaction vacuum state that is
revealed in the new theoretical frame-work.

 Before getting into the quantitative discussions, an important
qualitative property of the statistical blocking parameter
$\varepsilon^\mu$ should be mentioned. It is related to the
possibility of a distortion of the $\varepsilon^\mu$ around the
nucleon, which seems to happen at a first look since the statistical
gauge field $\mu^\alpha$ has distinct different values inside and
outside of a nucleon. Such a distortion does not happen for
$\varepsilon^\mu$.  Because the statistical gauge field $\mu^\alpha$
is coupled to a conserved current. A change of $\mu^\alpha$ in a
space-time region would cause a change of the baryon number density in
the same region. Such a change can not be balanced locally since
baryon number can not be created, it has to be transported inside or
outside of the region from other regions far apart which cause
corresponding changes of the baryon number in these other regions. In
the end, the statistical gauge field $\mu^\alpha$ correlates with each
other at long distances\footnote{Such a long distance correlation is
destroyed by the statistical blocking effects in the normal chiral
symmetry breaking phase or the $\alpha$-phase of the strong
interaction vacuum \cite{tp1,tp2,tp3,tp4}.}. The statistical blocking
parameter does not couple to a conserved number; it can
be seen as effectively couples to a current that is the difference of
the fermion's baryon number and that of the anti-fermion's baryon
number. This number can be created or destroyed locally (in space-time) in an
extremely efficient way since a generation or annihilation of one
fermion--anti-fermion pair inside of the vacuum would change two
unit of the charge of this later effective current. This means that
there is little correlation between $\varepsilon^\mu$ at different
space-time points. Therefore, $\varepsilon^\mu$ can be viewed as
corresponding to a very massive excitation of the system, which will
not response to the presence of a nucleon once its vacuum expectation
value is established.

\subsection{The vacuum statistical blocking parameter}

The absolute minima of the effective potential in the normal chiral
symmetry breaking phase are located at non-vanishing values of
$\varepsilon$ \cite{tp1,tp2,tp3,tp4} and vanishing $\mu^\alpha$. 
The effective potential is of
the following form
\begin{equation}
V_{\mbox{\scriptsize eff}}(\sigma,\varepsilon) = i 2n_f n_c \int_{{\cal
C}} {\frac{d^4p}{(2\pi)^4}} \ln \left ( 1-{\frac{\sigma^2
}{p^2}} \right ) + {\frac{1}{4 \widetilde G_0}} \sigma^2 + {n_f n_c\over
2\pi^2} \varepsilon^4,
\label{Veff3}
\end{equation}
where the dependency on the statistical gauge field $\mu^\alpha$ is
suppressed and the integration contour ${\cal C}$ lies beneath the
real $p^0$ axis for $-\infty < p^0 <-\varepsilon$ and
$0<p^0<\varepsilon$ and above the real $p^0$ axis when $-\varepsilon <
p^0 < 0 $ and $\varepsilon < p^0 < \infty$ \cite{tp1}. The minima for
$V_{\mbox{\scriptsize eff}}(\sigma,\varepsilon)$ are not located at
zero $\varepsilon$ for any non-vanishing $\sigma$.

For the case interested here, $\sigma=\sigma_{vac}=350$ MeV. The value
of the corresponding statistical blocking parameter
$\varepsilon_{vac}$ is found to be
\begin{equation}
  \varepsilon_{vac} \approx 40 \hspace{2mm} \mbox{MeV}
\end{equation} 
after minimizing of the effective potential.

\subsection{The energy of the valence quarks}

In the 8--component ``real'' theory for fermions, there are two
branches of single quark orbits with energy given by
\begin{equation}
  E_{\pm} = \pm \left (\epsilon_0 - \mu \right )
\end{equation}
that are relevant to the problem. Here $\epsilon_0$ is given by
Eq. \ref{eps0} and $\mu$ is given by \ref{muval}.  For each specific
value of $R$ value, only one of these two states corresponds to
particle excitation and the hole of the other state correspond to
mirror antiparticles \cite{ptct,Wsh}. The later excitation is absent
in the conventional 4-component theory for fermions. The other states
with 
\begin{equation}
   E'_\pm = \pm \left (\epsilon_0+\mu \right )
\end{equation}
lie in the continue spectra. They are not directly relevant to the
discussion here.

\begin{figure}[ht]
\begin{center}
\epsfbox{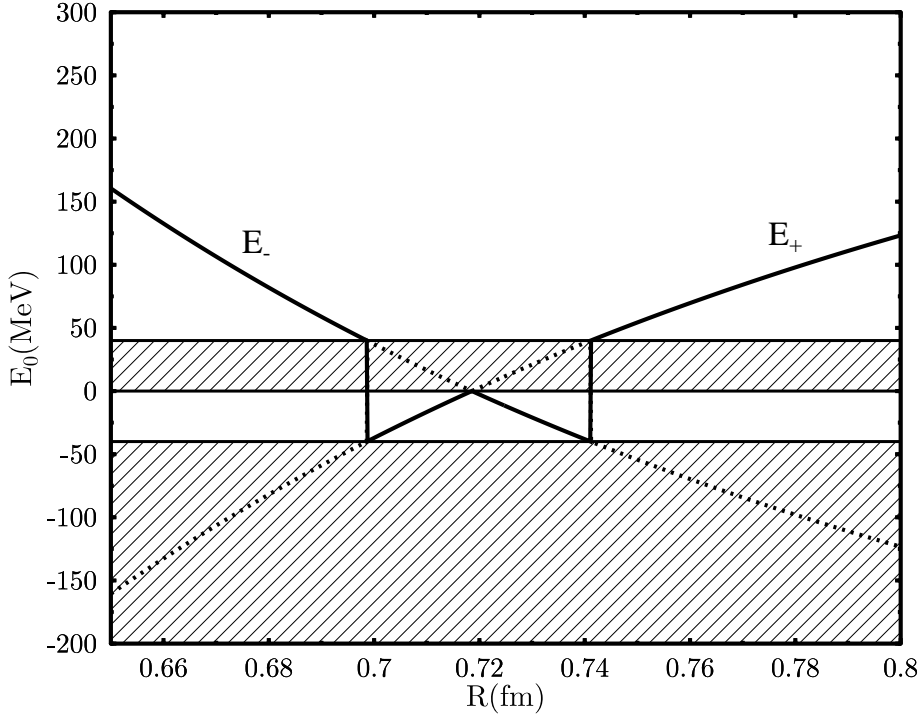}
\end{center}
\caption{\label{Fig:Edisc}\small\em The two branches of quark energy as
functions of $R$ in case of a constant $\Sigma_N(0)=62$ MeV. When
$R<0.65$ fm, $E_-$ is in the region of allowed state. The energy of
the quark orbit is given by $E_-$ in such a situation. As $R$ is
further increased, $E_-$ gets into the blocked states and $E_+$ state
becomes allowed. Such a switch of quark orbit release an energy of
order $80$ MeV for each quark. The role of $E_+$ and $E_-$ is switched
again at about $R\sim 0.72$ fm. It cause no energy in this case. The
last switch of $E_+$ and $E_-$ happens at $R\sim 0.72$ fm. An energy of
order $80$ MeV is needed to facilitate such a switch.}
\end{figure}

The allowed state in the presence of the statistical blocking is shown
in Fig. \ref{Fig:Edisc} for $\Sigma_N(0)=62$ MeV. The spectrum of the
particle excitation is draw with a solid line there. There are two
discontinuities for the particle energy level for the valence quarks,
with a gap of order of $80$ MeV. The rest of the curves draw in dotted
lines are for (mirror) anti-particles.
It is likely that the crossing of $E_+$ and $E_-$ at degenerate point
of about $R\sim 0.72$ fm is separated by a gap in more quantitative
studies.  Such a small complication is not going to be discussed
further here.  The important feature here is the presence of a gap for
the lowest energy of the quark orbits for bag radius $R\sim 0.70$ and $R\sim 0.74$
fm. These two gaps provide a stabilizing force for a nucleon.

\subsection{The ``quantization'' of the size of a nucleon}

The mass of a nucleon consists of two parts. The first one
is the energy of the valence quarks discussed above. The
second is the energy needed to establish the soliton configuration
in which the value of $\sigma$ is different from $\sigma_{vac}$ 
and there is a finite statistical gauge field $\mu^0$.
Combining these contributions, the ``classical value'' of the mass is given by
\begin{equation}
   M_N = 3 E_0(R) + {4\pi\over 3}B R^3 + {3\over
   2\pi^2}\Delta v \mu^4 - {1.5\over R},
\label{ClsMn}
\end{equation}
where $R$ is treated as classical quantity.  Here $E_0(R)$ is the
allowed energy for valence quarks, $B$ is the bag constant similar to
the MIT bag model and the third term is the corresponding term in the
local theory \cite{tp1}. The last term is a term from the free fermion
orbits corresponding to the valence quarks that has to be subtracted
according to the prescriptions of the local theory \cite{tp1}. There
should be a surface term that is proportional to $R^2$ in principle.
I shall ignore such a  term in the following discussions.

The value of $B$ can in principle be obtained from
$V_{\mbox{\scriptsize eff}}$ too. But it will not be fixed this way
here since the model for evaluating $V_{\mbox{\scriptsize eff}}$ does
not include many important elements such as the gluon degrees of
freedom, which may be important especially when $\sigma\ne
\sigma_{vac}$. It will be treated as free parameters here.  If the
value of $\Sigma_N(0)$ is chosen to be
\begin{equation}
 \Sigma_N(0) = 62 \hspace{2mm} \mbox{MeV}
\end{equation}
within the experimental range, then, it is found that $M_N\approx 1$
GeV if
\begin{equation}
   B = (219 \hspace{2mm} \mbox{MeV} )^4 = 2.3\times 10^{-3}
      \hspace{2mm}\mbox{GeV}^4.
\end{equation}

\begin{figure}[ht]
\begin{center}
\epsfbox{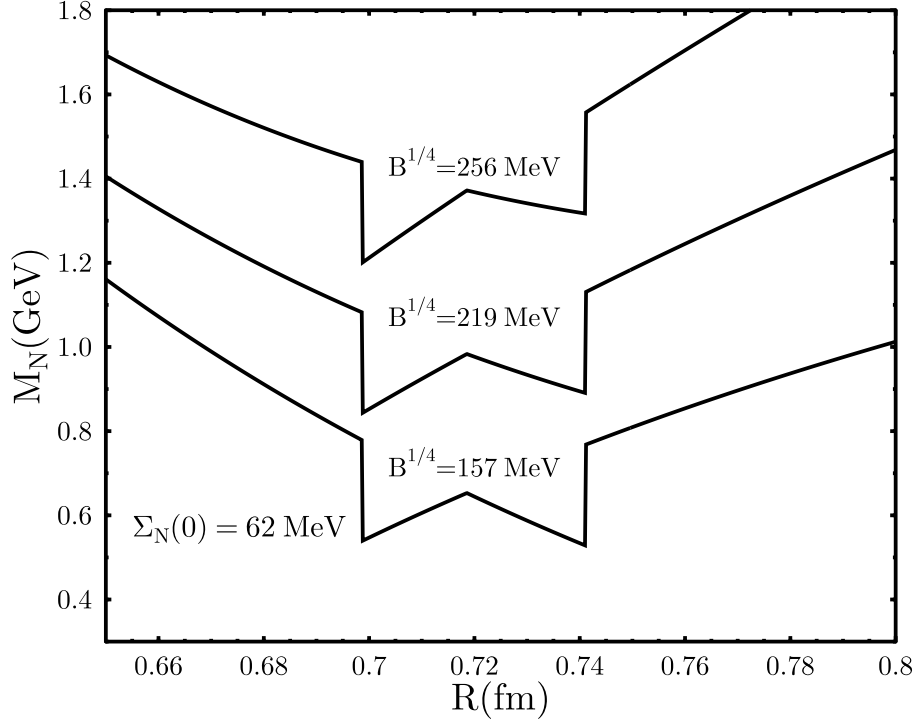}
\end{center}
\caption{\label{Fig:MnQn}\small\em The ``classical'' nucleon mass as a 
function of the
soliton radius $R$ for different values of the bag constant $B$.
The value of $\Sigma_N(0)$ is assumed independent of $R$. 
The radius for the soliton is fixed against the change of $B$ relative
to the value that reproduces the nucleon mass.
}
\end{figure}

The change of $M_N$ with $R$ is shown in Fig. \ref{Fig:MnQn}. It can
be seen that for the value of $\Sigma_N(0)$ chosen, the radius of a
nucleon is around $0.67\sim 0.78$ fm. Albeit there are two stable
positions for $M_N$ given in Eq. \ref{ClsMn}, a nucleon has a unique
radius due to the fact that so far $R$ is treated as a classical
variable. Since $R$ is a collective coordinate for a finite system
of confined quarks (including the sea quarks), it
has to be quantized, despite its fluctuation is expected to be much
reduced \footnote{If a collective coordinate, like $R$ considered
here, connects to the common motion of $N$ particles, the magnitude of
its quantum spreading is reduced by a factor of $1/\sqrt{N}$ compared
to the spreading of each of the single particles. }, it is
non-vanishing due to the finiteness of the system. If such a
quantization of $R$ is considered, the tunneling between these two
stable position leads to a unique stable value for $R$, which is in
the middle of the trapping well.

Also shown in
Fig. \ref{Fig:MnQn} are $M_N$ for different values of $B$. 
It can be seen that the location of the gap for $M_N$ is not changed
as $B$ is varied despite the significant change of its value. This is
an important property since one of the medium effects inside of a
nucleus is represented by the reduction of $B$ in the medium, which
causes the increase of the size of the nucleon in the conventional
picture. The statistical blocking effects in the local finite density
theory prevent such a change from happening.

In the authentic MIT model, the dependence of $M_N$ on the bag radius
$R$ is much flatter, which means that a nucleon is much softer than
the one studied here. For example, the size of the nucleon changes
rapidly from 0.75 fm to 1.1 fm as the bag constant $B^{1/4}$ is
reduced from $219$ MeV to $157$ MeV, which is already too large.

Due to the presence of a gap of order $240$ MeV for $M_N$, which is
three times of the gap for a single valence quark, a nucleon should
appear to be quite rigid against any form of external forces to
change its radial size. 

\section{The Local $\sigma_N$ Term and the Virtual Component of a
         Nucleon}
\label{sec:local}

  After establishing a possible stable chiral model for a nucleon in a
semi-quantitative way base on the phenomenological information about a
nucleon, we now turn to the local properties of the nucleon concerning
its possible color superconducting companion that lives in the
possible virtual color superconducting phase of the strong interaction
vacuum state.  Such a possibility seems to be supported by various
empirical evidences \cite{vcsc1,vcsc2,jpg,cmtp}. The results of these
works, however, depend only on general model independent properties of
the virtual color superconducting phase, like the spontaneous partial
breaking of the electromagnetic local gauge symmetry
\cite{vcsc1,vcsc2,jpg} and the spontaneous breaking of chiral symmetry
\cite{cmtp}. Detailed structure of this possible superconducting
component of a nucleon is not yet discussed.

   A few properties of the color superconducting component of a
nucleon that is likely to be true can nevertheless be listed. First,
the radius of such a component should be comparable to the normal
component discussed above. Second, there are also three valence
quarks, which is the quasi-particles of the color superconducting
phase, in such a component. This is because a nucleon is color
neutral. Third, the energy density of the observable color
superconducting phase(s) should not be much larger than the true
ground state of the vacuum. And fourth, the normal and the virtual
component of a nucleon should appear in pair in low energy processes.
Keeping this picture in mind, we shall turn to the question of the
static scalar charge of a nucleon.

   The nucleon $\Sigma_N$ problem is different from the leading order
relations because both side of the isoscalar part of the corresponding
chiral Ward--Takahashi identity for the nucleon $\Sigma_N$ problem are
sensitive to the low energy chiral dynamics.  The value for $\Sigma_N$
on the Cheng--Dashen point extracted from pion--nucleon scattering
data depends only on the pion nucleon coupling and, at low energy
interested here, it is a global observable.  The $\sigma_N$ obtained
from baryon spectrum contains the information of the possible
metastable color superconducting phase \cite{pcac1} if it indeed
exists. This is because it is a local operator in space-time defined
on an equal time hypersurface within the spatial region occupied by
the nucleon.  Therefore, according to the local theory
\cite{tp1,tp2,tp3,tp4} developed for such kind of situations, it contains
the dark component that is not seen in the global low energy
pion--nucleon scattering observables. It is therefore interesting to
see to what extent one can learn, from the difference between the
nucleon $\Sigma_N$ term extracted from the pion--nucleon scattering
data and the corresponding term extracted from the baryon mass
spectra, about the structure of the companion virtual superconducting
component of a nucleon that lives in the possible metastable color
superconducting phase of the strong interaction vacuum state and also
about the difference between the energy density of the true chiral
symmetry breaking phase and the virtual color superconducting phase of
the strong interaction vacuum.

\subsection{The static scalar charge of a nucleon as a local observable}
\label{sec:II}

The scalar charge of the nucleon that is extracted from the baryon spectrum
is the following expectation value of the
normal ordered scalar operator
\begin{equation}
    \sigma_N = m_0 \int d^3 x
                <N|:\overline\psi\psi:(\mbox{\boldmath{$x$}},t=0) |N>.
\label{sigm-def}
\end{equation}
The normal ordering is relative the vacuum state in which the chiral 
symmetry is spontaneously broken and it is measured at the
equal-time hypersurface of the rest frame of the nucleon. 

Although the detailed information about the right hand side (r.h.s)
of Eq. \ref{sigm-def} can not be easily extracted given the QCD dynamics,
some general properties of it can nevertheless be given. 
First, the matrix element $$<N|:\overline\psi\psi:(\mbox{\boldmath{$x$}},t) |N>$$
is non-vanishing only around the nucleon in its rest frame due to the normal
ordering. Therefore, the presence of a nucleon at the time $t=0$ as
specified in Eq. \ref{sigm-def} can be viewed as a local perturbation 
(measurement) of the vacuum state with spatial dimension of order
of the size of a nucleon and with a infinite resolution in time.
According to the local theory developed in Refs. \cite{tp1,tp2,tp3,tp4},
the quantity $\sigma_N$, which is a measure of certain component of the
{\em energy density} inside a nucleon, contains a dark component. 

 The dark component can be classified into two kinds. The first kind
of the dark component is from the deviation of the local observable
from the one that contains only the contribution of the
quasi--particles of the vacuum state. They are always present for
local observables.  For an semi-quantitative study given here, such a
component is relatively hard to evaluate and is expected to be
small. It is only included formally here. The second kind, which
is perhaps more interesting, is due to the possibility that there is a
close by metastable phase for the vacuum state. In this later case,
the quasi-particles from the metastable phase also contribute to the
local observables \cite{tp1,tp2,tp3,tp4} with a suppression factor ${\cal F}$
determined by the resolution of the observation $\Delta\omega$ and the
difference $\Delta\epsilon$ between the energy density  of the true
phase and the virtual phase(s) of the vacuum state.

The suppression factor can be determined according to the prescription 
given in Refs. \cite{tp1,tp2,tp3,tp4}. First, the quantum correlation length
for the observable interested is determined by the scale $1/M_A\sim 1$
$\mbox{GeV}^{-1}\sim 0.2$ fm, it is smaller than the size of the nucleon 
but is apparently larger than the inverse of the 
temporal resolution, which is zero. 
Therefore we have
\begin{equation}
{\cal F} = e^{-\Delta\epsilon \Delta v/M_A}
\end{equation}
for all one particle irreducible graphs constructed from the
quasiparticles of the virtual phase \cite{tp1}. Here $\Delta v$ is of
the order of the volume taken by a single nucleon in space.  Suppose
that there is a virtual color superconducting phase, then
Eq. \ref{sigm-def} can be further specified as
\begin{equation}
   \sigma_N = \sigma_{n} + \sigma_{s}
\end{equation}
where $\sigma_{n}$ is the contributions from the quasi-particles
(together with their dark component of the first kind discussed 
above) of the ground state of the vacuum in which the chiral symmetry
is spontaneously broken and $\sigma_s$ is the contributions from
the quasi-particles of the metastable virtual phase, which is
assumed to be color superconducting here, including all the 
dark component of the first kind. 

\subsection{The local $\sigma_N$ term}

For the local observables, the two type of dark components discussed
in section \ref{sec:II} has to be included. The dark components of the
first kind are included formally by doing the loop 4-momentum
integration in the Euclidean space which is cut off covariantly. The
dark component of the second kind, namely, the one coming from the
quasiparticles of the possible metastable phases of the vacuum, can be
evaluated using standard field theoretical method.
\begin{figure}[ht]
\begin{center}
\epsfbox{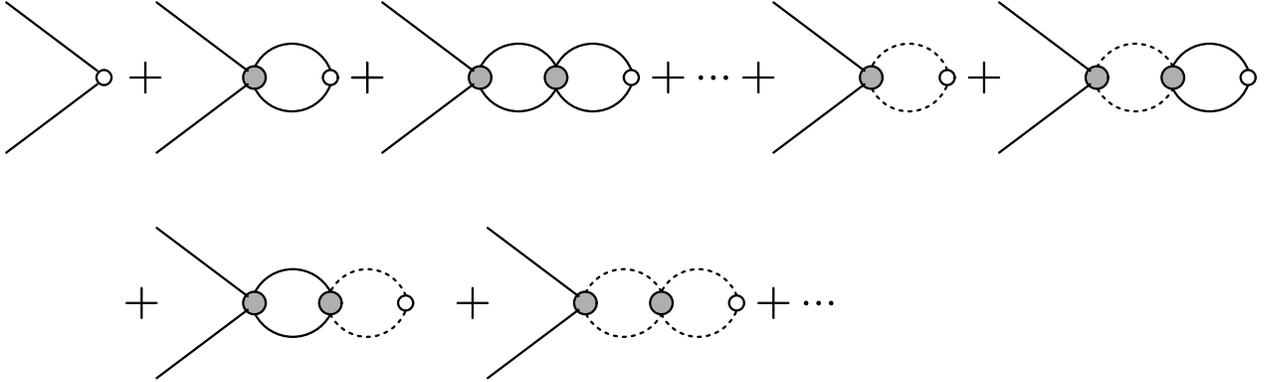}
\end{center}
\caption{\label{fig:scharge}\small\em The considered contribution to the local
scalar vertex function of a quark in a 4--fermion interaction model.
The lowest order graph in the upper row is the piece of the
static charge of the quark in the mean field approximation that is
usually included in the study of chiral symmetry breaking. The rest of
the higher order graphs are the contributing one loop graphs
in the conventional perturbation theories, which are represented by
solid lines and the non-perturbative one loop
contributions from the quasi-particles of the metastable phase whose
propagator is represented by a dash line.}
\end{figure}

The local scalar charge Eq. \ref{sigm-def}, can be written as
\begin{equation}
  \sigma_N = \Sigma_N + \delta\sigma_N^{(val)} + \delta\sigma_N^{(sea)}
\end{equation}
where $\Sigma_N$ is the global value discussed above,
$\delta\sigma_N^{(val)}$ is the contribution of the quasiparticles of
the metastable phase to the valence quarks and $\delta\sigma_N^{(sea)}$ 
is the contribution of the quasiparticles of
the metastable phase to the sea quarks.

  The type of the corrections to the scalar charge of a quark are be
diagrammatically expressed in  Fig. \ref{fig:scharge}.
Since the valence quarks of the color superconducting component of a
nucleon does not contain a scalar charge in the chiral limit. They do
not contribute to the quantities studied here. Therefore only the
Feynman graphs containing external quark lines living in the normal
phase of the vacuum state are draw in Fig. \ref{fig:scharge}. 

After the summation of the series of the ladder diagrams and including
the wave function renormalization for the quark legs, the correction
to the scalar charge of a quark is found to be
\begin{equation}
   \delta j_s =  - J_s \overline u(p) u(p) 
\end{equation}
with the one loop diagram contribution denoted by $J_s$. The value of
$J_s$ depends on the nature of the color superconducting phase.

\subsubsection{The possible virtual scalar color superconducting
   phase}

The propagators for the quarks in the scalar color superconducting
phase can be found after fixing the complex
phase and the direction in the color space of $\chi^c$. If $\chi^c$
is chosen to be along, say, the ``red'' direction and to be real, then
\cite{sdiq,tp1}
\begin{equation}
 S_{F1}(p) = {i\over \rlap\slash p - \gamma^5 {\cal A} \chi O_1}
\end{equation}
for ``blue'' and ``green'' quarks and
\begin{equation}
 S_{F2}(p) = {i\over \rlap\slash p}
\end{equation}
for ``red'' quark.

The bubble diagram in Fig. \ref{fig:scharge} is found to be
\begin{eqnarray}
 J_s &=& -i\widetilde G_0 {\cal F}\mbox{Tr} \int {d^4p\over (2\pi)^4} \left [
            S_{F1}(p)S_{F1}(p) + S_{F2}(p)S_{F2}(p) \right ] 
 \nonumber \\
    &=& 16 \widetilde G_0 {\cal F}\int {d^4p_E\over (2\pi)^4} \left (
      {2\over p^2_E + \chi^2} + {1\over p_E^2}
\right )
\end{eqnarray}
with $p_E^\mu$ the Euclidean 4-momentum. The result is
\begin{equation}
  J_s = {12\alpha_0\over \pi}{\cal F}\left [1-{2\over 3}
    {\chi^2\over\Lambda^2} \ln\left (1+{\Lambda^2\over\chi^2} \right )
\right ]  = {\cal F} {1-{\displaystyle 2\over\displaystyle 3}
    {\displaystyle\chi^2\over\displaystyle\Lambda^2} 
     \ln \left (1+{\displaystyle\Lambda^2\over\displaystyle\chi^2}
     \right )\over
     1-{\displaystyle\sigma^2\over\displaystyle\Lambda^2} 
     \ln \left (1+{\displaystyle\Lambda^2\over\displaystyle\sigma^2}
    \right )}.
\end{equation}
Here, use has been made of the gap equation for $\sigma$, namely Eq.
\ref{gap-eq}.  Since $\sigma\approx 350$ MeV and assuming $\chi\sim
200$ MeV, then $J_s=1.3\times e^{-\Delta\epsilon \Delta v/M_A}$.

The contribution of the quasiparticles of the virtual phase to the
value of the local $\sigma_N(0)$ can be obtained from Eq. \ref{Sig1}
through a replacement of the global wave function renormalization
factor $1-J_n$ by local one, namely $1-J_n \to 1-J_n-J_s$. Since the wave
function renormalization factor for every contributing quark orbits,
including the sea quarks, is modified in such a way, we can write
\begin{equation}
  \sigma_N(0) = \Sigma_N(0) - {J_s\over 1-J_n}\Sigma_N(0).
\end{equation}
 Therefore we have
\begin{equation}
   J_s = {(1-J_n)\Delta\Sigma_N\over \Sigma_N(0) }.
\end{equation}
For the value of $\chi$ taken, we have
\begin{equation}
  \Delta \epsilon = {3\over 4\pi} {M_A\over R^3} \ln {1.3 \times\Sigma_N(0)\over
  (1-J_n)\Delta\Sigma_N(0)}. 
\label{Deleps1}
\end{equation}
Using the estimates $\Sigma_N(2\mu^2)-\Sigma_N(0)=10\sim 15$ MeV from
the chiral perturbation theory \cite{chpt} and dispersion relation
calculation \cite{dispe}, $\Sigma_N(0) = 55\sim 75$ MeV. If no
strangeness is assumed for a nucleon and $M_A=\Lambda=0.9$ GeV, then
\begin{equation}
  \Delta\epsilon = 789 \sim 409 \hspace{2mm} \mbox{MeV/fm$^3$}.
\label{Delepsval}
\end{equation}
This is a reasonable number.

\subsubsection{The possible virtual vector color superconducting phase}

The propagator represented by the dashed lines in
Fig. \ref{fig:scharge} for the vector color superconducting phase are
also found after fixing the color and complex phase for the order
parameter $\phi_\mu^c$.  One of the choices \cite{Ann1} is that for
``blue'' and ``green'' quarks
\begin{eqnarray}
   S_{F1}(p) &=& \left (1-iO_2{\rlap\slash p\over p^2}\rlap\slash\phi
               \gamma^5{\cal A} \right ) F(p),\\
   F(p) &=& i{(p^2-\phi^2)\rlap\slash p-2p\cdot\phi \rlap\slash \phi\over
            (p^2-\phi^2)^2-4(p\cdot\phi)^2}
\end{eqnarray}
and
\begin{equation}
 S_{F2}(p) = {i\over \rlap\slash p}
\end{equation}
for ``red'' quark.

The bubble diagram in Fig. \ref{fig:scharge} is found to be
\begin{eqnarray}
 J_s &=& -i\widetilde G_0 \mbox{Tr} \int {d^4p\over (2\pi)^4} \left [
            S_{F1}(p)S_{F1}(p) + S_{F2}(p)S_{F2}(p) \right ] 
 \nonumber \\
    &=& 16 \widetilde G_0 \int {d^4p_E\over (2\pi)^4} \left (
      {2(p^2_E+\phi^2)\over (p^2_E + \phi^2)-4(p_E\cdot\phi)^2} + {1\over p_E^2}
\right )
\end{eqnarray}
with $p_E^\mu$ the Euclidean 4-momentum. The result is quite simple
\begin{equation}
  J_s = {12\alpha_0\over \pi}{\cal F}\left (1-{1\over 3}{\phi^2\over\Lambda^2}
        \right ) = 
     {\cal F} {1-{\displaystyle 1\over\displaystyle 3}
    {\displaystyle\phi^2\over\displaystyle\Lambda^2} 
     \over
     1-{\displaystyle\sigma^2\over\displaystyle\Lambda^2} 
     \ln \left (1+{\displaystyle\Lambda^2\over\displaystyle\sigma^2}
    \right )}.
\end{equation}

For typical values of $\sigma_{vac}=350$ MeV and $\sqrt{\phi^2}=200$ MeV, 
$J_s=1.4\times e^{-\Delta\epsilon
\Delta v/M_A}$ and Eq. \ref{Deleps1} is replaced by
\begin{equation}
  \Delta \epsilon = {3\over 4\pi} {M_A\over R^3} \ln {1.4
  \times\Sigma_N(0)\over (1-J_n)\Delta\Sigma_N(0)},
\label{Deleps2}
\end{equation}
which is not much different from Eq. \ref{Deleps1}. The values for
$\Delta\epsilon$ is of the same order as Eq. \ref{Delepsval}, namely
\begin{equation}
  \Delta\epsilon = 853 \sim 450  \hspace{2mm} \mbox{MeV/fm$^3$}.
\label{Delepsval2}
\end{equation}
This is a reasonable number. 

This range of $\Delta\epsilon$ is quite stable against change of the
order parameters for the color superconducting phase. For example, even
in the limit $\sqrt{\phi^2}\to 0$, 
\begin{equation}
  \Delta\epsilon = 865 \sim 457  \hspace{2mm} \mbox{MeV/fm$^3$},
\label{Delepsval3}
\end{equation}
which changes very little. The same is true for the scalar color
superconducting phase.

\section{Discussions}
\label{sec:discussion}

\subsection{The inputs and outputs}

The input parameters are listed in Table \ref{Tab:I}.
\begin{table}[ht]
\caption{\small\em The input parameters and equations. The value of
$J_n$ is consistent with the recent lattice evaluation. The value of
$c_0$ is a prediction of the MIT bag model. It is treated as an input
to reflect our ignorance of the shape of $\sigma(r)$. \label{Tab:I}}
\begin{center}
\begin{tabular}
{|c@{\hspace{0.3in}}c@{\hspace{0.3in}}c@{\hspace{0.3in}}
c@{\hspace{0.3in}}c@{\hspace{0.3in}}c@{\hspace{0.3in}}c@{\hspace{0.3in}}
          c@{\hspace{0.3in}}|}
\hline
\hline
&&&&&&&\\
 & $\alpha_0$ & $\Lambda$ & $J_n$ & $\chi$ or $\sqrt{\phi^2}$ & 
 B &$c_0$ & \\
&&&&&&&\\
 & 0.38 & 0.9 GeV  & 0.21 & 0.2 GeV &  $2.3\times 10^{-3}$ GeV$^4$ & 2.04 &
  \\
&&&&&&&\\
\hline\hline
\end{tabular}
\end{center}
\end{table}
The quantitative outputs of our investigation based on the local
theory are given in Table \ref{Tab:II}. It can be seen that using 6
input parameters/equations, more than double physical
quantities/equations can be obtained/satisfied. The gain is
obvious. In addition, a new mechanism for the stability of a nucleon
inside a nucleus or nuclear matter is found and semi-quantitative
pictures for the nucleon and the vacuum state of the strong
interaction is also established.
\begin{table}[ht]
\caption{\label{Tab:II}\small\em The physical/empirical values generated by the
model in the local theory. These values are not fitted. They are
derived values according to the local theory.  The known quantities
agree well with the popular ones in literature.  The new parameters of
the local theory, namely $\mu$, $\Delta\epsilon$, $\varepsilon_{vac}$
and the stability gap, are predicted in this work.  All these
quantities are consistent with the chiral symmetry encoded in the
Ward--Takahashi (WT) identities. The number in bracket is the observed
one. $M^*_N$ correspond to the case of $\Sigma_N(0)=62$ MeV. The unit
for various quantities here is defined the text.}
\begin{center}
\begin{tabular}{|c|cccccc|}
\hline\hline
&&&&&&\\
Physical & $f_\pi$ & $M^*_N$ & $\Sigma_N(0)$ & $\sigma_N(0)$ & 
$<0|\overline\psi\psi |0>$ & \\
&&&&&&\\
Quantities &  $\begin{array}{c} 92\\ (93) \end{array}$  & 
$\begin{array}{c}1.0\\ (0.938) \end{array}$ 
& $\begin{array}{c} 55 \sim 75\\
   (55\sim 75) \end{array}$ & 35 (35) 
 & $\begin{array}{c} -0.024\\ (-0.024)\end{array}$ & \\
&&&&&&\\
\hline
&&&&&&\\
Physical & $\sigma_{vac}$ &  $R$  & $\sigma_{in}$ &
$\mu$ & 
$\delta\hspace{-1mm}<0|\overline\psi\psi |0>$ &   \\
&&&&&&\\
Quantities &   350  & $0.67\sim 0.78$  & $-0.15\sim -0.13$  & 
0.45$\sim$ 0.38  & 0.034$\sim$ 0.032 &   \\
&&&&&&\\ 
\hline
&&&&&&\\
Physical & $\Delta\epsilon_s$ (Scalar) & $\Delta\epsilon_v$ (Vector) & 
$\varepsilon_{vac}$ & Stability gap & &\\
&&&&&&\\
Quantities       & $0.79\sim 0.41$  & $0.85\sim 0.45$ 
&40   & $\sim 240$  && \\
&&&&&&\\
\hline
&&&&&&\\
Chiral Symmetry & Eq. \ref{GORrel} &
    Eq. \ref{fpi} & $\Sigma_N$ puzzle & Nucleon stability&  &\\
&&&&&&\\
\& Others & Satisfied  & Satisfied  & Solved & Ensured  & &\\
&&&&&&\\
\hline\hline
\end{tabular}
\end{center}
\end{table}

 The presence of at least one metastable virtual color superconducting
phase for the strong interaction vacuum is assumed instead of derived
in this paper. Unlike in other work, like Refs. \cite{jpg,vcsc1,cmtp},
the existence of the metastable color superconducting phase for the
strong interaction vacuum state is a sufficient condition for the
solution of the nucleon $\Sigma_N$ term puzzle. It is not a necessary
one despite the fact that such an assumption is rather natural for
the solution of the puzzle. 

\subsection{The nucleon as a non-topological soliton}

It is demonstrated that when the quark wave function renormalization
is taken into account, the experimentally observed global nucleon
$\Sigma_N$ term implies, in general, that a modification of the
quasi--particle picture for the hadrons seems to be
inevitable.  This kind of picture for a nucleon is somewhat different
from the authentic constituent quark model but are in qualitative
agreement with the bag type of models \cite{mit,flee,Cloudy1,Cloudy2}
for a nucleon introduced on more phenomenological basis where the
nature of the scalar $\sigma$ field, which is claimed to be related to
gluon contributions, is actually not clear. Here the scalar field is
the order parameter for the chiral symmetry, which contains
contributions from gluon as well as quarks. The sizes of the
non-topological soliton are found to be around $0.7$ fm.

Given the fact that $\sigma_{vac}$, the bag constant $B$ and the
renormalization factor $J_n$ are related to the complicated underlying
QCD dynamics, they are not computed using the 4-fermion interaction
models adopted here, which serve the purpose of providing the general
structure of the solution to the problems and of discussing the chiral
symmetry issues.  Rather, these parameters are determined by fitting
empirical data.  This is done despite the fact that model computations
may not leads to a set of values significantly deviate from the fitted
ones, such a success may not be regarded as an indication that QCD
formulated in the local theory can be replaced by the models but only
that these models are effective ones.

\subsection{The solution of the two puzzles}

After the basic picture of a nucleon is established, the role played
by the statistical blocking effects of the chiral symmetry breaking
phase in maintaining the stability of a nucleon inside a large nucleus
and in a nuclear matter is discussed. It is shown that two of the
essential elements of the local theory, namely, the statistical
blocking effects and the inequivalent 8-component ``real''
representation for the fermion field are both needed for the
mechanism. The bag constant $B$ is determined in the frame-work of
the local theory by producing the right mass for the nucleon.

The discrepancy between the nucleon $\Sigma_N$ term from low energy
pion--nucleon scattering data and the one extracted from baryon
spectra is resolved by realizing that the former one is a global
observable while the later one is a local one, which according to the
local theory, contains a dark component. The dark component is further
attributed to a possible virtual color superconducting phase that
seems to manifest in the high energy electromagnetic interactions
involving a nucleon \cite{vcsc1,vcsc2,jpg} and some other ones
\cite{review}. The difference in energy density of the metastable
phase and the stable phase of the strong interaction vacuum state is
estimated.

  Thus, the new features of the local theory \cite{tp1,tp2,tp3,tp4} and
the assumption that there is a virtual color superconducting phase for
the strong interaction vacuum state provides a set of conditions for a
simultaneous solution of the two puzzles.

\subsection{The diquark model for a nucleon?}

 The diquark models for a nucleon has a long history. It has certain
advantages on the phenomenological ground. There are theoretical
difficulties for these models. It is criticized in a similar, albeit
more detailed in certain aspects, approach to the quark--quark
interaction, which does not consider the required wave function
renormalization \cite{iter}. We belief that their argument is sound
at least for this specific problem.

But under the current scenario, the result of Ref. \cite{iter} is not
the end of the diquark model for a nucleon since its argument applies
only to the normal component of the nucleon, which lives in the true
ground state of the vacuum in which the chiral symmetry is
spontaneously broken. In the case of the possible virtual color
superconducting component for a nucleon, which lives in the possible
virtual color superconducting phase of the vacuum, an entirely
different set of channels, like in the $\chi$ or $\phi^\mu$,
$\mbox{\boldmath{$\delta$}}^\mu$ etc. fields must be iterated. It is
not known at the present whether or not a quark--quark clustering for
the ``constituent'' quarks in the color superconducting component of a
nucleon is favored or not. It is one of the questions to be understood
in the future.

\subsection{Speculations}

Given these figures, it is tempting to speculate that perhaps the
Ropper resonance at $1.44$ GeV is an excited $N^*$ resonance produced
in such a way that its normal component is striped away by some means
so that it is dominated by the color superconducting companion of a
nucleon? It would be very interesting to understand the structure of
the color superconducting virtual companion of the nucleon and more
importantly the virtual color superconducting phase of the strong
interaction vacuum state if they are indeed there.

\section{Summary}
\label{sec:summary}
 
   A mechanism for the stability of a nucleon and a resolution of the
nucleon $\Sigma_N$ term problem are studied based on an implementation
of the chiral Ward--Takahashi identity of QCD at a level that goes
beyond the mean field approximation by formally summing over ladder
diagrams for $\sigma'$ field. The local theory for the problem
\cite{tp1,tp2,tp3,tp4} is adopted as the theoretical frame-work for a
coherent discussion and solution of these two seemingly unrelated
puzzles.

In the future, self-consistent numerical study of the non-topological
soliton model proposed here is an interesting topic to be studied
\cite{soliton}. Theoretical and experimental means should be searched
and/or developed to study the structure of the possible
superconducting component of the nucleon in more details. It is
interesting not only to the nucleon structure itself, but also because
it carries information about the possible virtual color
superconducting phase of the vacuum state, which is one of the most
important objects to be understood in the context of contemporary 
discoveries in cosmology.

\section*{Acknowledgement}
 
This work is supported by the National Natural Science Foundation of China
under contract 19875009. Part of the present work was carried out while
the author was visiting the CSSM of the University of Adelaide, Adelaide,
Australia.


\end{document}